\documentclass[journal,twoside,web]{ieeecolor}
\usepackage{tmi2}
\usepackage{cite}
\usepackage{amsmath,amssymb,amsfonts}
\usepackage{graphicx}
\usepackage{textcomp}
\usepackage{verbatim}
\usepackage{booktabs}
\usepackage{tabularray}
\usepackage{mathtools} 
\usepackage{algpseudocode}
\usepackage[ruled,vlined]{algorithm2e}
\SetKwInput{KwInput}{Input}                
\SetKwInput{KwOutput}{Output}              

\def\BibTeX{{\rm B\kern-.05em{\sc i\kern-.025em b}\kern-.08em
    T\kern-.1667em\lower.7ex\hbox{E}\kern-.125emX}}

\newcommand{\TtwoCT}{T\textsubscript{2}$\rightarrow$CT}
\newcommand{\ToneCT}{T\textsubscript{1}$\rightarrow$CT}
\newcommand{\Tone}{T\textsubscript{1}}
\newcommand{\Ttwo}{T\textsubscript{2}}

\newcommand{\ToneTtwo}{T\textsubscript{1}$\rightarrow$T\textsubscript{2}}
\newcommand{\TonePD}{T\textsubscript{1}$\rightarrow$PD}
\newcommand{\TtwoTone}{T\textsubscript{2}$\rightarrow$T\textsubscript{1}}

\newcommand{\PDTone}{PD$\rightarrow$T\textsubscript{1}}

\newcommand{\TtwoFlair}{T\textsubscript{2}$\rightarrow$FLAIR}
\newcommand{\FlairTtwo}{FLAIR$\rightarrow$T\textsubscript{2}}

\usepackage{tikz}
\usetikzlibrary{matrix,calc}
\usepackage{framed,multirow}
\usepackage{latexsym}
\usepackage{url}
\usepackage{xcolor}
\usepackage{lipsum}
\usepackage{hyperref}
\hypersetup{
    colorlinks=true,
    linkcolor=blue,
    filecolor=magenta,      
    urlcolor=cyan,
    pdftitle={SelfRDB},
    }
\usepackage{gensymb}
\usepackage{caption}
\usepackage{comment}

\definecolor{newcolor}{rgb}{.8,.349,.1}
\usepackage[math]{cellspace}
\makeatletter
\IEEEtriggercmd{\reset@font\normalfont\fontsize{7.9pt}{8.40pt}\selectfont}
\makeatother
\IEEEtriggeratref{1}

\begin{document}
\title{Self-Consistent Recursive Diffusion Bridge\\ for Medical Image Translation}
\author{Fuat Arslan, Bilal Kabas, Onat Dalmaz, Muzaffer Ozbey, and Tolga \c{C}ukur$^\ast$ 
\vspace{-0.9cm}
\\
\thanks{This study was supported in part by TUBA GEBIP 2015 and BAGEP 2017 fellowships awarded to T. \c{C}ukur. F. Arslan and B. Kabas contributed equally to the study. (Corresponding author: Tolga \c{C}ukur, cukur@ee.bilkent.edu.tr)}
\thanks{F. Arslan, B. Kabas and T. \c{C}ukur are with the Dept. of Electrical-Electronics Engineering and National Magnetic Resonance Research Center (UMRAM), Bilkent University, Ankara, Turkey, 06800. T. \c{C}ukur is also with the Dept. of Neuroscience, Bilkent University, Ankara, Turkey, 06800. O. Dalmaz is with the Dept. of Electrical Engineering, Stanford University, CA 94305. M. Ozbey is with the Dept. of Electrical and Computer Engineering, University of Illinois Urbana-Champaign,  IL 61820.}
}

\maketitle
\begin{abstract}
Denoising diffusion models (DDM) have gained recent traction in medical image translation given improved training stability over adversarial models. DDMs learn a multi-step denoising transformation to progressively map random Gaussian-noise images onto target-modality images, while receiving stationary guidance from source-modality images. As this denoising transformation diverges significantly from the task-relevant source-to-target transformation, DDMs can suffer from weak source-modality guidance. Here, we propose a novel self-consistent recursive diffusion bridge (SelfRDB) for improved performance in medical image translation. Unlike DDMs, SelfRDB employs a novel forward process with start- and end-points defined based on target and source images, respectively. Intermediate image samples across the process are expressed via a normal distribution with mean taken as a convex combination of start-end points, and variance from additive noise. Unlike regular diffusion bridges that prescribe zero variance at start-end points and high variance at mid-point of the process, we propose a novel noise scheduling with monotonically increasing variance towards the end-point in order to boost generalization performance and facilitate information transfer between the two modalities. To further enhance sampling accuracy in each reverse step, we propose a novel sampling procedure where the network recursively generates a transient-estimate of the target image until convergence onto a self-consistent solution. Comprehensive analyses in multi-contrast MRI and MRI-CT translation indicate that SelfRDB offers superior performance against competing methods.
\end{abstract}
\begin{IEEEkeywords}
medical image translation, synthesis, generative, diffusion, bridge  \vspace{-0.1cm}
\end{IEEEkeywords}

\bstctlcite{IEEEexample:BSTcontrol}

\section{Introduction}
Medical images acquired via multiple modalities capture complementary diagnostic information on bodily tissues \cite{iglesias2013,lee2017}, but running multi-modal protocols is burdening given associated economic and labor costs  \cite{ye2013,huynh2015,jog2017,joyce2017}. A powerful framework to extend the exam scope without elevating costs is medical image translation, wherein unacquired target modalities are predicted from acquired source modalities \cite{cordier2016,wu2016,zhao2017,huang2018}. Important clinical applications of translation include imputation of target modalities to lower protocol redundancy or to avoid harmful contrast agents/ionizing radiation \cite{lee2019}, and facilitating participation in retrospective imaging studies by extending the protocol scope and homogeneity across participants \cite{divbar2019}. That said, medical image translation is a heavily ill-posed problem as signal levels for a given tissue show nonlinear variations across modalities \cite{roy2013,alexander2014,huang2017}. As such, learning-based methods that excel at inverse problems have recently become the de facto framework for medical image translation \cite{hien2015,vemulapalli2015,sevetlidis2016,nie2016}. 

\begin{figure*}[t]
\centering
\includegraphics[width=\linewidth]{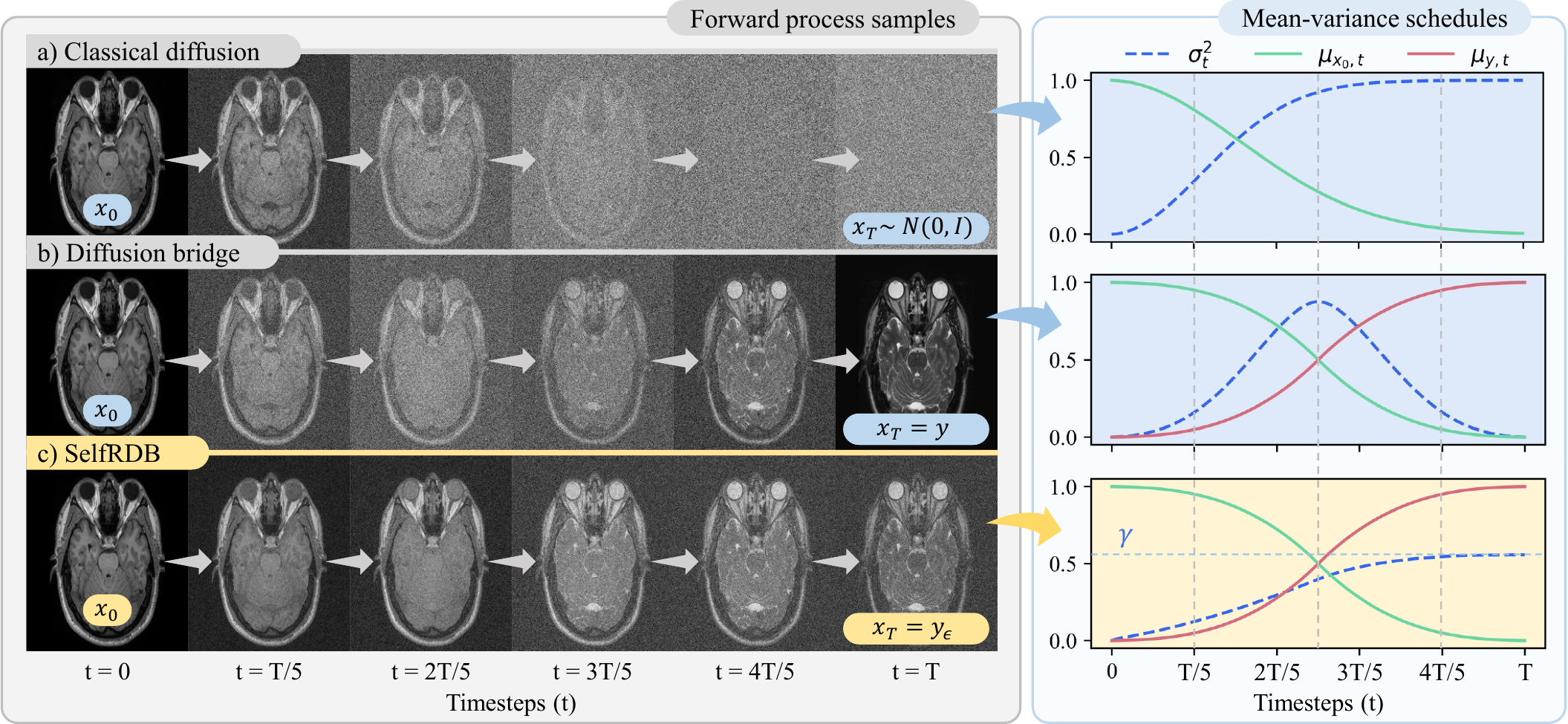}
\caption{Diffusion methods commonly take the target image as the start-point $\boldsymbol{x}_0$ of the diffusion process, albeit they can differ in expression of image samples in remaining timesteps. Illustrations of images across the forward process are depicted along with underlying schedules for the mean ($\mu_{x_0,t}$, $\mu_{y,t}$) and noise variance ($\sigma_{t}^2$). \textbf{(a) Classical diffusion:} DDMs use a pure noise image as an asymptotic end-point $\boldsymbol{x}_T$. Intermediate samples are obtained by adding increasing levels of random Gaussian noise onto the target image. \textbf{(b) Diffusion bridge:} Regular bridges use the source image as a finite end-point. Intermediate samples are taken as a convex combination of source-target images, corrupted with additive noise. Noise variance is zero at start- and end-points, and it peaks at the mid-point. \textbf{(c) Proposed:} SelfRDB is a novel diffusion bridge that uses a noise-added source image as the end-point. Intermediate samples still depend on a convex combination of source-target images, yet SelfRDB uniquely prescribes monotonically-increasing noise variance towards the end-point.}
\label{fig:forwards}
\end{figure*}

Learning-based methods commonly aim to capture a conditional prior for target given source images, albeit differ in the way that they learn this prior \cite{bowles2016,chartsias2018,wei2019}. Among previous methods, generative adversarial networks (GAN) have been widely adopted for their exceptional realism in synthesized images \cite{yu2018,armanious2019,li2019}, and successfully reported in diverse tasks including translation between MR contrasts and MRI-CT \cite{nie2018,dar2019image,yu2019,yang2018,jin2018,resvit}. Yet, since GANs capture an implicit prior through a generator-discriminator interplay, they can suffer from training instabilities that hamper image fidelity \cite{wang2020,zhou2020}. To improve stability, recent studies have employed denoising diffusion models (DDM) to instead capture an explicit prior \cite{syndiff,meng_arxiv_2022,lyu_arxiv_2022,wang2024tmi}. In DDMs, a forward process gradually degrades the target image by repeated addition of Gaussian noise till an asymptotic end-point of pure noise (Fig. \ref{fig:forwards}a). Starting with the random noise image, a reverse process then progressively denoises the input via a network to recover the target image, while the source image provides stationary guidance \cite{syndiff}. Despite their stability, DDMs learn a task-irrelevant denoising transformation from noise to target images, which can weaken source-image guidance \cite{song2021solving,adadiff}. In turn, DDMs can perform suboptimally in medical image translation given the divergence between the learned denoising transformation and the required source-to-target transformation \cite{DDPM,adadiff}.

\begin{figure*}[t]
    \centering
    \includegraphics[width=\linewidth]{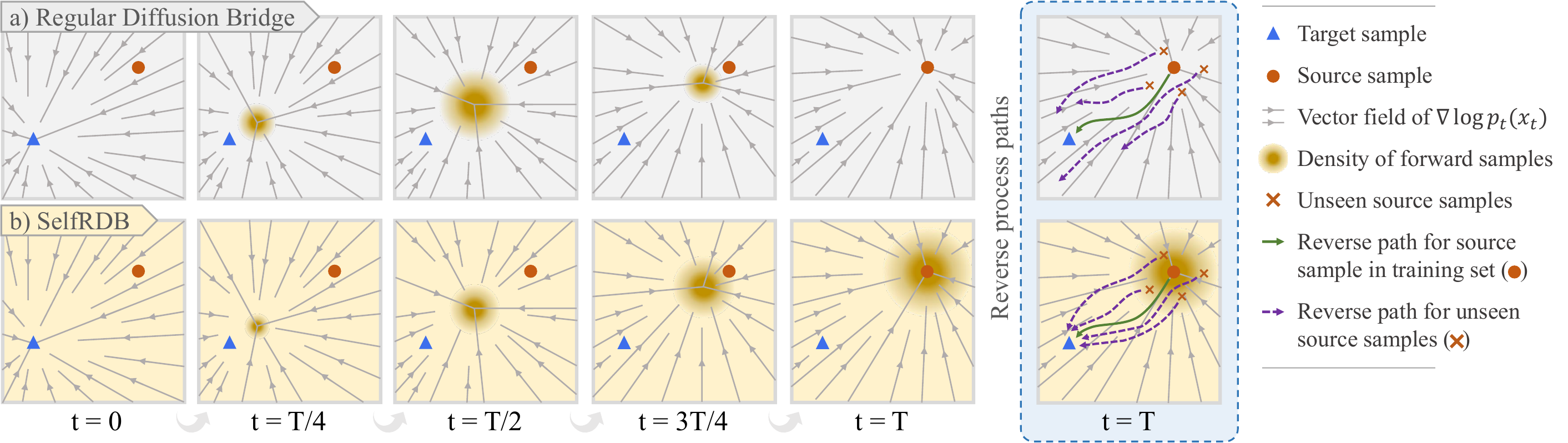}
    \caption{Diffusion models learn the score function of the data through a multi-step transformation between the start- and end-points of the underlying process. Image samples are typically corrupted with Gaussian noise that smooths the data distribution by masking some of the original image features. Smoothing enables more uniform coverage of the data space in order to boost generalization performance. \textbf{(a) Regular diffusion bridges} use zero noise variance at the end-point constraining them to a Dirac-delta distribution centered on the source images within the training set. This can compromise generalization performance to source images outside the training set (see purple-colored dashed paths). \textbf{(b) SelfRDB} instead uses monotonically-increasing variance towards the end-point, so it is trained on noise-added source images. This improves robustness against variability in source images between training and test sets (see purple-colored dashed paths).}
    \vspace{-0.2cm}
    \label{fig:soft_prior}
\end{figure*}

An emerging approach to enhance task relevance in diffusion-based priors employs diffusion bridges that can directly transform between two separate modalities \cite{delbracio2023inversion,chung2023direct}. To do this, diffusion bridges define the start- and end-points of the forward process based on target and source images, respectively (Fig. \ref{fig:forwards}b). As the imaging operator linking the two modalities is typically unknown, image samples in intermediate steps are derived from a normal distribution whose mean is a convex combination of start- and end-points \cite{liu2023i2sb,kim2024unpaired}. Initiating sampling on the source image, the reverse process progressively maps the source onto the target image. Few recent imaging studies have successfully employed diffusion bridges in the reconstruction of single-modal images from undersampled or low-resolution measurements \cite{FDB,kim2024hicbridge}. However, the potential of diffusion bridges in medical image translation remains largely unexplored, as existing methods face several key challenges. Regular diffusion bridges adopt a noise scheduling with zero variance at start-end points albeit high variance near the mid-point of the diffusion process \cite{DDIB}. Zero variance at the end-point results in \textit{a hard-prior on the source modality} reflecting a Dirac-delta distribution centered on source images within the training set, hampering generalization (Fig. \ref{fig:soft_prior}a). Meanwhile, heavy noise at the mid-point can disrupt source-to-target information transfer. Furthermore, diffusion bridges typically synthesize a \textit{one-shot estimate of intermediate samples}, limiting sampling accuracy \cite{peng2022}.

\begin{figure*}[t]
\centering
{\includegraphics[width=0.7\linewidth]{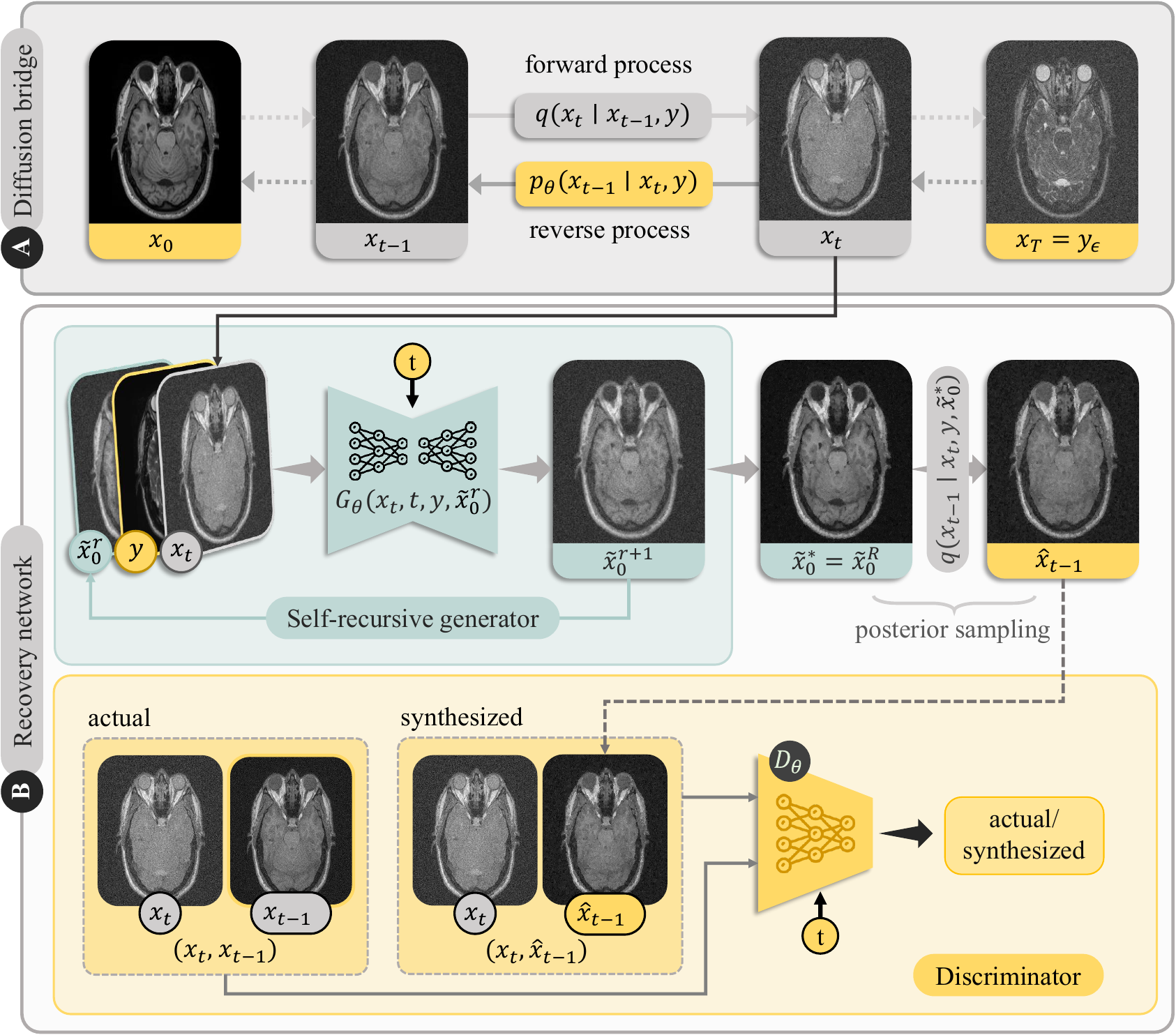}}
\caption{SelfRDB casts a diffusion bridge between source and target images of an anatomy. \textbf{(a)} In the forward process, the start-point $\boldsymbol{x}_0$ is taken as the target image and the end-point $\boldsymbol{x}_T$ is taken as a noise-added version of the source image $\boldsymbol{y}_{\epsilon}$. Intermediate image samples are derived via the forward transition probability $q(\boldsymbol{x}_t | \boldsymbol{x}_{t-1},\boldsymbol{y})$, whose mean is a convex combination of target-source images, and whose variance is driven by noise. In the reverse process, sampling is initiated on $\boldsymbol{x}_T=\boldsymbol{y}_{\epsilon}$, and intermediate samples are derived via the reverse transition probability $p_{\theta}(\boldsymbol{x}_{t-1} | \boldsymbol{x}_{t},\boldsymbol{y}_{\epsilon})$. \textbf{(b)} Reverse diffusion steps are operationalized via a recovery network $G_{\theta}(\boldsymbol{x}_t,t,\boldsymbol{y},\tilde{\boldsymbol{x}}_0^r)$ that recursively generates a target-image estimate $\tilde{\boldsymbol{x}}_0^{r+1}$ at the current timestep, given the target-image estimate from the previous recursion $\tilde{\boldsymbol{x}}_0^{r}$ and the original source image $\boldsymbol{y}$. Recursions are stopped upon convergence onto a self-consistent solution $\tilde{\boldsymbol{x}}_0^*=G_{\theta}(\boldsymbol{x}_t,t,\boldsymbol{y},\tilde{\boldsymbol{x}}_0^*)$, which is then used for posterior sampling of $\hat{\boldsymbol{x}}_{t-1}$ according to the normal distribution $q(\boldsymbol{x}_{t-1}|\boldsymbol{x}_t,\boldsymbol{y},\tilde{\boldsymbol{x}}_0^*)$. To improve posterior sampling, a discriminator subnetwork $D_{\theta}(\boldsymbol{x}_{t-1} \mbox{ or } \hat{\boldsymbol{x}}_{t-1},t,\boldsymbol{x}_t)$ is used to perform adversarial learning on the recovered sample $\hat{\boldsymbol{x}}_{t-1}$.}
\label{fig:method}
\end{figure*}

Here, we propose a novel self-consistent recursive diffusion bridge, SelfRDB, to improve performance in multi-modal medical image translation. Unlike regular diffusion bridges, SelfRDB leverages a novel noise scheduling in its forward process, with monotonically increasing variance towards the end-point that corresponds to a noise-added source image (Fig. \ref{fig:forwards}c). As such, it captures \textit{a soft-prior on the source modality} to attain improved generalization, while it facilitates information transfer between modalities by prescribing lower noise near the mid-point of the process (Fig. \ref{fig:soft_prior}b). To avoid loss of tissue information at the noise-added end-point, SelfRDB's recovery network employs stationary guidance from the original source image in the reverse process. Finally, to improve sampling accuracy in each reverse step, SelfRDB leverages a novel \textit{self-consistent recursive estimation procedure} for the target image, and uses this self-consistent estimate to synthesize intermediate samples with enhanced accuracy (Fig. \ref{fig:method}). Comprehensive demonstrations are performed for multi-contrast MRI and MRI-CT translation. Our results clearly indicate the superiority of SelfRDB against competing GAN and diffusion models, including previous diffusion bridges. Code for SelfRDB is available at \href{https://github.com/icon-lab/SelfRDB}{https://github.com/icon-lab/SelfRDB}. 

\vspace{-0.25cm}
\subsection*{Contributions}
\begin{itemize}
\item To our knowledge, SelfRDB is the first diffusion bridge for medical image translation between separate modalities in the literature. 
\item SelfRDB leverages a novel forward diffusion process that captures a soft-prior on the source modality to improve generalization and facilitate information transfer between source-target modalities. 
\item SelfRDB leverages a novel self-consistent recursive estimation procedure to improve sampling accuracy in reverse diffusion steps. 

\end{itemize}

\section{Related Work}
DDMs have recently been adopted in multi-modal medical image translation given their improved image fidelity \cite{syndiff,meng_arxiv_2022,lyu_arxiv_2022,wang2024tmi}.
Employing a forward process where target images are corrupted with additive noise over a large number of steps, DDMs progressively map a random noise image onto the target under stationary guidance from the source image \cite{DDPM}. This multi-step denoising transformation helps improve training stability over GANs \cite{wolterink2017,dong2019}. Unfortunately, the image mapping performed by the denoising transformation is weakly associated with the desired source-to-target image mapping for translation tasks, and the source-image guidance in DDMs is primarily implicit \cite{liu2023i2sb}. In turn, these problems can compromise performance in DDM-based translation. Here, we introduce the first diffusion bridge for multi-modal medical image translation to our knowledge. Unlike DDMs that express intermediate samples as noise-added target images and use an end-point of Gaussian noise, SelfRDB expresses intermediate samples as a convex combination of source and target images corrupted with additive noise, and employs an end-point of a noise-added source image. Unlike DDMs that use one-shot sampling in each reverse step, SelfRDB employs self-consistent recursive estimation to improve sampling accuracy. Based on these unique advances, we provide the first demonstrations of multi-contrast MRI and MRI-CT translation based on diffusion bridges in the literature.  


Diffusion bridges are an emerging alternative to DDMs to improve flexibility in generative modeling tasks. Several computer vision studies \cite{daras2022soft,delbracio2023inversion,chung2023direct} and a few recent imaging studies \cite{FDB,kim2024hicbridge} have devised diffusion bridges for single-modality reconstruction tasks, with the aim to recover an image from linearly corrupted measurements (e.g., blurred, undersampled or low-resolution). Unlike single-modal diffusion bridges, SelfRDB performs a translation task to map between distinct source and target modalities whose relationship is uncharacterized. Several recent computer vision studies have also built diffusion bridges for multi-modal translation tasks \cite{DDIB,liu2023i2sb,kim2024unpaired}. However, regular diffusion bridges were commonly employed based on a noise schedule with zero variance at start-end points corresponding to target-source images, yet high variance near the mid-point. This scheduling can hamper generalization to source images in the test set  \cite{chung2023direct}, and induce substantial masking of tissue information near the mid-point during source-to-target mapping \cite{liu2023i2sb}. To address these limitations, SelfRDB uniquely leverages a monotonically-increasing noise variance towards the end-point. Furthermore, compared to previous single- and multi-modal bridges that use a one-shot sampling process in reverse steps, SelfRDB leverages a novel self-consistent recursive estimation procedure to improve accuracy in generation of intermediate image samples. 

\section{Theory and Methods}

\subsection{Diffusion Bridges}
Diffusion bridges are a general framework to describe the evolution between two arbitrary probability distributions across a finite time interval $t \in [0,T]$ \cite{liu2023i2sb}. In the context of mapping a source image $\boldsymbol{x}_T:=\boldsymbol{y}$ onto a target image $\boldsymbol{x}_0$, the learning objective for diffusion bridges can be expressed as:
\begin{equation}
\min _{p \in \mathcal{P}_{[0,T]}} D_\text{KL}(p \| q), \quad \text{s.t.} \; p_0=p_{\text {target}}, p_T=p_{\text {source},}
\label{eq:sb_kl}
\end{equation}
where $\mathcal{P}_{[0,T]}$ is the space of path measures with marginal densities for the target and source ($p_0=p_{\text {target }}$ and $p_T=p_{\text {source}}$) taken as boundary conditions, and $q$ is the reference path measure. Solution of \eqref{eq:sb_kl} is the optimal path measure $p^* \in \mathcal{P}_{[0,T]}$ that can be described via the following forward-reverse stochastic differential equations \cite{chen2021likelihood}:
\begin{align}
\mathrm{d}\boldsymbol{x}_t &= [\boldsymbol{f} + g^2 \nabla \log \Psi(\boldsymbol{x}_t, t)]\mathrm{d}t + g \mathrm{d}\boldsymbol{w}_t, \: \boldsymbol{x}_0\sim p_{\text{target}}, \nonumber \\ 
\mathrm{d}\boldsymbol{x}_t &= [\boldsymbol{f} - g^2 \nabla \log \bar{\Psi}(\boldsymbol{x}_t, t)]\mathrm{d}t + g {\mathrm{d}\bar{\boldsymbol{w}}_t}, \: \boldsymbol{x}_T\sim p_{\text{source}}. \label{eq:SDE}
\end{align}
Here, $\boldsymbol{f}$ is the drift coefficient, $g$ is the diffusion coefficient, $\boldsymbol{w}_t, \bar{\boldsymbol{w}}_t$ are forward-reverse Wiener processes, and $\nabla \log \Psi(\boldsymbol{x}_t, t)$, $\nabla \log \bar{\Psi}(\boldsymbol{x}_t, t)$ are nonlinear forward-reverse drift terms related to the score function $\nabla \log p_t(\boldsymbol{x}_t)$ \cite{nelson1967}. In contrast to DDMs based on linear drifts \cite{song2020score}, the nonlinear drifts in diffusion bridges enable the use of non-Gaussian $p_{\text{source}}$. A Dirac-delta distribution is assumed for the target modality, i.e., $p_0(\cdot) := \delta_x (\cdot)$, such that the marginal density at the start-point is taken as $p_0(\boldsymbol{x}_0) = 1$ given a (target, source) image pair $(\boldsymbol{x}_0, \boldsymbol{x}_T)$. This assumption ensures computational tractability by decoupling the constraints in Eq. \eqref{eq:SDE} \cite{liu2023i2sb}.

In regular diffusion bridges, high-quality image pairs from target and source modalities are taken as start- and end-points of the diffusion process as in Eq. \eqref{eq:sb_kl}, with zero additive-noise variance at $t=0$ and $t=T$ (Fig. \ref{fig:forwards}b). This choice ensures optimal transport for training data, and enables the bridge to directly translate high-quality source images during inference \cite{liu2023i2sb}. However, it also constrains the bridge to capture \textit{a hard-prior on the source modality}, since the end-point follows a Dirac-delta distribution based on source images in the training set, i.e., $p_T(\boldsymbol{x}_T)=1$ given a training pair $(\boldsymbol{x}_0, \boldsymbol{x}_T)$. Combined with high noise variance near $t=T/2$, the distributional constraint compromises generalization and information transfer from source-to-target images (Fig. \ref{fig:soft_prior}a). 

\subsection{SelfRDB}
SelfRDB is a novel diffusion bridge for medical image translation that maps the source image $\boldsymbol{y} \sim p_\text{source}$ of an anatomy onto the respective target image $\boldsymbol{x}_0 \sim p_\text{target}$. To do this, it leverages a novel forward process with a soft-prior on the source modality to improve generalization and facilitate information transfer, and a novel reverse process with self-consistent recursion to improve sampling accuracy.

\vspace{0.2cm}
\subsubsection{Forward process with soft-prior on source modality} SelfRDB forms a diffusion bridge between $\boldsymbol{x}_0$ and $\boldsymbol{y}$ based on the following forward transition probability (Fig. \ref{fig:forwards}):
\begin{equation}
    q(\boldsymbol{x}_t| \boldsymbol{x}_0, \boldsymbol{x}_T) = \mathcal{N}(\boldsymbol{x}_t; \mu_{x_0,t} \boldsymbol{x}_0 + \mu_{y,t} \boldsymbol{y}, \sigma_t^2 \boldsymbol{I}), \label{eq:q_sample}
\end{equation}
where $\boldsymbol{x}_t$ is the intermediate image sample at timestep $t$, $\mathcal{N}$ denotes the Gaussian distribution, and $\boldsymbol{I}$ is the identity matrix. Accordingly, given an image pair ($\boldsymbol{x}_0$, $\boldsymbol{y}$), intermediate image samples are generated as follows:
\begin{equation}
\boldsymbol{x}_t = \mu_{x_0,t} \boldsymbol{x}_0 + \mu_{y,t} \boldsymbol{y} + \sigma_t \boldsymbol{\epsilon}, \label{eq:genforward}
\end{equation}
where $\boldsymbol{\epsilon} \sim \mathcal{N}(\boldsymbol{0},\boldsymbol{I})$ is a standard normal variable. Note that the mean of $\boldsymbol{x}_t$ is determined via a convex combination of target and source images with weights $\mu_{x_0,t}$, $\mu_{y,t}$ \cite{liu2023i2sb}:
\begin{gather}
    \mu_{x_0,t} = \frac{\bar{s}_t^2}{\bar{s}_t^2 + s_t^2}, \quad 
    \mu_{y,t} = \frac{s_t^2}{\bar{s}_t^2 + s_t^2}, \label{eq:meanschedule}
\end{gather}
where $s_t^2 \coloneqq \int_0^t g(\tau) d\tau$ and $\bar{s}_t^2 \coloneqq \int_t^T g(\tau) d\tau$ are the time-accumulated diffusion coefficients in forward and reverse directions. To satisfy positivity and symmetry conditions with respect to the mid-point \cite{chen2023schrodinger}, here we propose to use the following diffusion coefficient:

\begin{equation}
g(t) \propto \frac{(T-|2 t-T|)^2}{4 T(T-1)^2}.
\label{eq:diff_coeff}
\end{equation}

Meanwhile, the variance of $\boldsymbol{x}_t$ depends on the scale parameter $\sigma_t^2$. In regular diffusion bridges, $\sigma_t^2$ is defined to follow Dirac-delta constraints at start- and end-points \cite{liu2023i2sb}:
\begin{gather}
    \sigma^2_{t} = \frac{\bar{s}_t^2 s_t^2}{\bar{s}_t^2 + s_t^2} \text{ (regular bridge)},
\end{gather}
where $\sigma^2_{t}$ peaks at $t=T/2$ and is reduced to $0$ at $t=T$ resulting in an end-point $\boldsymbol{x}_T=\boldsymbol{y} \sim \mathcal{N}(\boldsymbol{x}_T; \boldsymbol{y}, \boldsymbol{0})$. Regular bridges learn an exact mapping between a target image and its paired source image, resulting in a hard-prior on the source modality. This can compromise generalization during inference on a source image drawn from a low-density region of the data space poorly covered in the training set \cite{song2019generative}.  

In contrast, SelfRDB adopts a novel noise variance schedule where $\sigma^2_{t}$ grows monotonically across $t$:
\begin{equation}
    \sigma^2_{t} =\gamma \frac{s_t}{\bar{s}_t^2 + s_t^2} \text{ (SelfRDB),} 
\end{equation}
where $\gamma$ is a scalar. Note that $\sigma^2_{t}$ ranges in $[0 \mbox{ } \gamma]$, so $\gamma$ is tuned as a hyperparameter to control the level of noise corruption at the end-point. The above schedule elicits an end-point of a noise-added source image $\boldsymbol{x}_T=\boldsymbol{y}_{\epsilon} \sim \mathcal{N}(\boldsymbol{x}_T;\boldsymbol{y},\sigma^2_T \boldsymbol{I})$. Noise addition relaxes the Dirac-delta constraint on the data distribution at the end-point, and smooths the corresponding data space to enable more uniform coverage. In turn, SelfRDB learns a mapping between a target image and the neighborhood of its paired source image (Fig. \ref{fig:soft_prior}). The resultant soft-prior serves to enhance the reliability of SelfRDB against variability in source images, thereby boosting generalization performance.     

\vspace{0.25cm}
\subsubsection{Reverse process with self-consistent recursion}
 SelfRDB casts a reverse process to progressively map the noise-added source image at the end-point $\boldsymbol{x}_T=\boldsymbol{y}_{\epsilon}$ back onto the target image $\boldsymbol{x}_0$ at the start-point (see Alg. \ref{alg:sampling}). Since $\boldsymbol{y}_{\epsilon}$ is corrupted by additive noise, stationary guidance from the original source image $\boldsymbol{y}$ is also employed to avoid losses in tissue information. Starting sampling at $\boldsymbol{x}_T$, intermediate image samples are drawn based on a network operationalization of the reverse transition probability $p_{\theta}(\boldsymbol{x}_{t-1}|\boldsymbol{x}_t,\boldsymbol{y})$$:=$$q(\boldsymbol{x}_{t-1}|\boldsymbol{x}_t,\boldsymbol{y})$. An adversarial recovery network with parameters $\theta$ is adopted here to learn $p_{\theta}(\boldsymbol{x}_{t-1}|\boldsymbol{x}_t,\boldsymbol{y})$ as inspired by the recent success of adversarial diffusion models in synthesis tasks \cite{DiffNvidia,syndiff}. 
 
 Assuming that the diffusion process comprises a sufficiently large number steps (i.e., $T$$\gg$$1$), $p_{\theta}(\boldsymbol{x}_{t-1}|\boldsymbol{x}_t,\boldsymbol{y})$ can be analytically expressed by reparametrizing the reverse transition probability as $q(\boldsymbol{x}_{t-1}|\boldsymbol{x}_{t},\boldsymbol{y},\boldsymbol{x}_0)$ \cite{syndiff}. However, the actual $\boldsymbol{x}_0$ is unknown at timestep $t$ during reverse diffusion. Thus, SelfRDB employs a generator $G_{\theta}$ to produce a target-image estimate $\tilde{\boldsymbol{x}}_0^*$ as a surrogate of $\boldsymbol{x}_0$. In an individual reverse step, diffusion methods commonly produce a one-shot image estimate by performing a single forward-pass through the recovery network \cite{DDPM}. Yet, deviations in synthesized image samples (i.e., $\boldsymbol{x}_t$) from the true data distribution can accumulate across reverse diffusion steps, causing significant estimation errors \cite{peng2022}. Unlike previous diffusion methods, SelfRDB leverages a novel self-consistent recursive estimation procedure to improve sampling accuracy: 
\begin{equation}
    \tilde{\boldsymbol{x}}_0^{r+1} = G_{\theta}(\boldsymbol{x}_t,t,\boldsymbol{y},\tilde{\boldsymbol{x}}_0^{r}),
\end{equation}
where $r \in \mathbb{Z}^{+}$ denotes the recursion index, and $\tilde{\boldsymbol{x}}_0^{r}$ is the target-image estimate at the $r$th recursion. Initially setting $\tilde{\boldsymbol{x}}_0^{1}=\boldsymbol{0}$, recursions are continued until a self-consistent solution $\tilde{\boldsymbol{x}}_0^*$ is obtained at the $R$th recursion:
\begin{equation}
    \tilde{\boldsymbol{x}}_0^* = G_{\theta}(\boldsymbol{x}_t,t,\boldsymbol{y},\tilde{\boldsymbol{x}}_0^*),
\end{equation}
This recursive estimation procedure gives a chance for the generator to correct intermittent estimation errors across recursions, thereby improving the accuracy of target-image estimates. 

\begin{algorithm}[t]
    \caption{Inference for SelfRDB}
    \label{alg:sampling}
\KwIn{\\ $\boldsymbol{y}$: original source image, $\boldsymbol{y}_{\epsilon}$: noise-added source image \\ 
$G_{\theta}(\boldsymbol{x}_t,t,\boldsymbol{y},\tilde{\boldsymbol{x}}_0)$: recovery network \\ 
$T$: number of diffusion steps \\
$R$: number of recursions \\} 
\KwOut{\\ $\hat{\boldsymbol{x}}_0$: recovered target image  \\ {$\mbox{ }$}}
 \vspace{-0.25cm} 
 $\boldsymbol{x}_T=\boldsymbol{y}_{\epsilon}$ \quad $\triangleright$ set end-point sample \\
         \For{$t = T, \hdots, 1$} {
             $\tilde{\boldsymbol{x}}^1_0 = \boldsymbol{0}$ \quad $\triangleright$ initialize target-image estimate \\
             \For{$r = 1, \hdots, R$} {
                 $\tilde{\boldsymbol{x}}^{r+1}_0 = G_{\theta}(\boldsymbol{x}_t, t, \boldsymbol{y}, \tilde{\boldsymbol{x}}_0^{r})$ \quad $\triangleright$ update estimate \\
             }
             $\tilde{\boldsymbol{x}}^*_0 = \tilde{\boldsymbol{x}}^R_0$ \quad $\triangleright$ retrieve self-consistent estimate \\
             $\hat{\boldsymbol{x}}_{t-1} \sim q(\boldsymbol{x}_{t-1}|\boldsymbol{x}_t,\boldsymbol{y},\tilde{\boldsymbol{x}}^*_0)$ \quad $\triangleright$ posterior sampling
        }
 \KwRet{$\hat{\boldsymbol{x}}_0$}
\end{algorithm}

Once an accurate target-image estimate $\tilde{\boldsymbol{x}}_0^*$ is derived, the image sample at timestep $t-1$ can be drawn from the reparametrized posterior by taking $\tilde{\boldsymbol{x}}_0^*$ as a surrogate for ${\boldsymbol{x}}_0$: 
\begin{equation}
    \hat{\boldsymbol{x}}_{t-1} \sim q(\boldsymbol{x}_{t-1}|\boldsymbol{x}_{t},\boldsymbol{y},\tilde{\boldsymbol{x}}_0^*)
\end{equation}
Based on Bayes' rule and Markov property of the diffusion process \cite{DDPM}, the posterior can be expressed as:
\begin{equation}
 q(\boldsymbol{x}_{t-1}|\boldsymbol{x}_{t},\boldsymbol{y},\tilde{\boldsymbol{x}}_0^*) = \frac{q(\boldsymbol{x}_{t}|\boldsymbol{x}_{t-1},\boldsymbol{y}) q(\boldsymbol{x}_{t-1}|\boldsymbol{y},\tilde{\boldsymbol{x}}_0^*)}{q(\boldsymbol{x}_{t}|\boldsymbol{y},\tilde{\boldsymbol{x}}_0^*)}. \label{eq:decomp}
\end{equation}
Note that, in Eq. \ref{eq:decomp}, the terms in the fractional expression can be computed based on the forward transition probability in Eq. \ref{eq:q_sample}. In turn, here we derive the posterior probability as a Gaussian distribution $\mathcal{N}(\boldsymbol{x}_{t-1};\boldsymbol{m},\boldsymbol{v})$ such that:
\begin{align}
\boldsymbol{m} = &\frac{\sigma_{t-1}^2}{\sigma_{t}^2} \frac{\mu_{x_0,t}}{ \mu_{x_0,t-1}} \boldsymbol{x}_t +
(\mu_{y,t-1}-\mu_{y,t}\frac{\sigma_{t-1}^2}{\sigma_{t}^2} \frac{\mu_{x_0,t}}{ \mu_{x_0,t-1}}) \boldsymbol{y} \nonumber \\+
& (1- \mu_{y,t-1}\frac{\sigma_{t|t-1}^2}{\sigma_{t}^2}) \tilde{\boldsymbol{x}}_0^*, \\
\boldsymbol{v} = & \sigma_{t|t-1}^2\frac{\sigma_{t-1}^2}{\sigma_{t}^2},
\end{align}
where $\sigma_{t|t-1}^2 = \sigma_{t}^2 - \sigma_{t-1}^2 (\mu_{x_0,t}/\mu_{x_0,t-1})^2$. 

The recovery network also employs a discriminator $D_\theta$ to distinguish the synthetic samples produced with the aid of the generator from the actual image samples drawn using the forward diffusion process. Conditioned on $\boldsymbol{x}_t$, $D_\theta$ predicts a logit of the input sample at timestep $t-1$:
\begin{equation}
c = D_\theta((\hat{\boldsymbol{x}}_{t-1} \text{ or } \boldsymbol{x}_{t-1}), t, \boldsymbol{x}_{t}).
\end{equation}

\vspace{0.2cm}
\subsubsection{Learning procedure}
Given a training set of target-source image pairs ($\boldsymbol{x}_0,\boldsymbol{y}$), the forward process described in Eqs. \ref{eq:q_sample}-\ref{eq:genforward} is used to generate corresponding intermediate samples $\boldsymbol{x}_t$ for $t \in [0 \mbox{ } T]$ that bridge between each image pair. Afterward, these intermediate samples are used to train the adversarial recovery network in SelfRDB. The generator aims to produce accurate target-image estimates $\tilde{\boldsymbol{x}}_0^*$ that subsequently elicit realistic intermediate image samples $\hat{\boldsymbol{x}}_{t-1}$. Thus, $G_\theta$ is trained using pixel-wise $\ell_1$ and adversarial loss terms \cite{pgan}:
\begin{eqnarray}
    L_{G_{\theta}} = \mathbb{E}_{t, q(\boldsymbol{x}_t|\boldsymbol{x}_0,\boldsymbol{y}),p_{\theta}(\boldsymbol{x}_{t-1}|\boldsymbol{x}_{t},\boldsymbol{y})} \big\{ \lambda_1 \| \boldsymbol{x}_0 - \tilde{\boldsymbol{x}}_0^* \|_1 \notag \\
     -log(D_{\boldsymbol{\theta}}(\hat{\boldsymbol{x}}_{t-1})) \big\},
\end{eqnarray}
where $\mathbb{E}$ is expectation, $\lambda_1$ is the weight of the pixel-wise loss. Meanwhile, the discriminator primarily aims to distinguish between synthetic and actual intermediate image samples, so $D_\theta$ is trained an adversarial loss with a gradient penalty \cite{pgan}:
\begin{eqnarray}
    L_{D_{\theta}} = \mathbb{E}_{t, q(\boldsymbol{x}_t|\boldsymbol{x}_0,\boldsymbol{y})} \Big\{ \mathbb{E}_{q(\boldsymbol{x}_{t-1}|\boldsymbol{x}_{t},\boldsymbol{y})} \big\{ -log(D_{\boldsymbol{\theta}}({\boldsymbol{x}}_{t-1})) \big\} \nonumber \\
    + \mathbb{E}_{p_{\theta}(\boldsymbol{x}_{t-1}|\boldsymbol{x}_{t},\boldsymbol{y})} \big\{ -log(1-D_{\boldsymbol{\theta}}(\hat{\boldsymbol{x}}_{t-1})) \big\} \nonumber \\
    + \lambda_2 \mathbb{E}_{q(\boldsymbol{x}_{t-1}|\boldsymbol{x}_{t},\boldsymbol{y})} \big\{ \| \nabla{\boldsymbol{x}_{t-1}} D_{\boldsymbol{\theta}}({\boldsymbol{x}}_{t-1}) \|^2_2 \big\} \Big\},
\end{eqnarray}
where $\lambda_2$ is the relative weight of the gradient penalty \cite{syndiff}.

\section{Experiments}

\subsection{Datasets}  
Experiments were conducted on two multi-contrast MRI datasets (IXI\footnote{https://brain-development.org/ixi-dataset/}, BRATS \cite{brats_1}) and a multi-modal MRI-CT dataset \cite{mr_ct_dataset}. In each dataset, a three-way split was performed to create training, validation and test sets with no subject overlap. Separate volumes of a subject were spatially registered via affine transformation \cite{fslcitation}. Each volume was normalized to a mean intensity of 1, and voxel intensities were then normalized to a range of [-1,1] across subjects. A consistent 256$\times$256 cross-sectional image size was attained via zero-padding.

\subsubsection{IXI Dataset} \Tone-, \Ttwo-, PD-weighted brain images from $40$ healthy subjects, with ($25$,$5$,$10$) subjects reserved for (training,validation,test), were analyzed. In each volume, $100$ axial cross-sections with brain tissue were selected.

\subsubsection{BRATS Dataset} \Tone-, \Ttwo-, Fluid Attenuation Inversion Recovery (FLAIR) weighted brain images from $55$ glioma patients, with ($25$,$10$,$20$) subjects reserved for (training,validation,test) were analyzed. In each volume, $100$ axial cross-sections containing brain tissue were selected.

\subsubsection{MRI-CT Dataset}
\Tone-, \Ttwo-weighted MRI, and CT images of the pelvis from $15$ subjects, with ($9$,$2$,$4$) subjects reserved for (training,validation,test) were analyzed. In each volume, $90$ axial cross-sections were selected. 

\begin{figure*}
    \centering
    \includegraphics[width=0.9\textwidth]{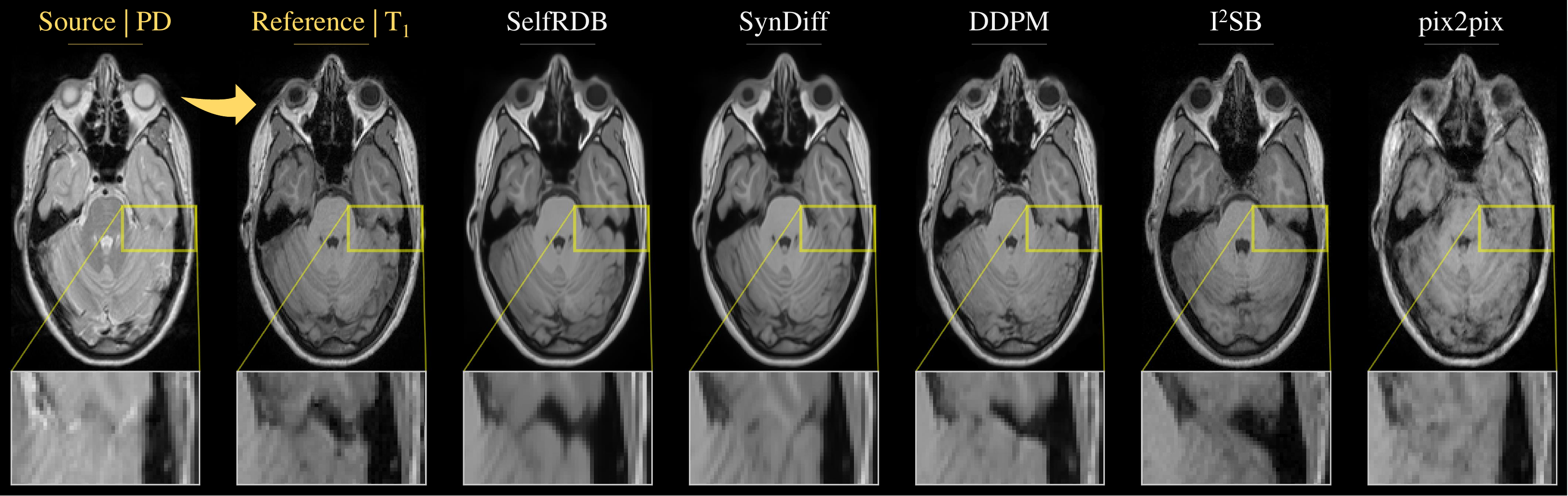}
    \caption{Multi-contrast MRI translation for a representative PD$\rightarrow$T\textsubscript{1} task in the IXI dataset. Synthesized target images for competing methods are shown along with the reference target image (i.e., ground truth) and the input source image. Zoom-in display windows are used to highlight differences in synthesis performance.}
    \label{fig:ixi_samples}
\end{figure*}

\begin{table*}[t]
\caption{Multi-contrast MRI translation in IXI. PSNR (dB) and SSIM (\%) are listed as mean$\pm$std across the test set. Boldface marks the top-performing model in each task.}
\centering

\begin{tabular}{lcccccccc}
\hline
\multirow{2}{*}{} & \multicolumn{2}{c}{\TtwoTone} & \multicolumn{2}{c}{\ToneTtwo}  & \multicolumn{2}{c}{\PDTone} & \multicolumn{2}{c}{\TonePD} \\ 
\cline{2-9} & PSNR & SSIM & PSNR & SSIM & PSNR & SSIM & PSNR & SSIM \\ \hline

\multirow{1}{*}{SelfRDB} 
& \textbf{31.56$\pm$1.58} & \textbf{95.65$\pm$1.18} 
& \textbf{30.70$\pm$1.53} & \textbf{94.90$\pm$1.29} 
& \textbf{31.05$\pm$1.27} & \textbf{95.69$\pm$0.99}
& \textbf{31.93$\pm$1.53} & \textbf{95.18$\pm$0.97} 
\\ \hline

\multirow{1}{*}{SynDiff} 
& 30.13$\pm$1.38 & 94.60$\pm$1.23
& 30.19$\pm$1.45 & 94.24$\pm$1.36
& 29.74$\pm$1.34 & 94.81$\pm$1.12
& 30.89$\pm$1.42 & 94.20$\pm$1.04
\\ \hline

\multirow{1}{*}{DDPM} 
& 29.08$\pm$1.05 & 93.43$\pm$1.35
& 29.89$\pm$1.29 & 94.31$\pm$1.33
& 29.98$\pm$0.89 & 94.45$\pm$1.04
& 30.58$\pm$1.27 & 93.76$\pm$1.08
\\ \hline

\multirow{1}{*}{I\textsuperscript{2}SB} 
& 21.07$\pm$0.47 & 47.06$\pm$1.85
& 21.98$\pm$0.55 & 77.61$\pm$1.90
& 21.61$\pm$0.42 & 77.95$\pm$1.81
& 24.88$\pm$0.80 & 79.44$\pm$1.94
\\ \hline

\multirow{1}{*}{pix2pix} 
& 27.98$\pm$1.06& 92.18$\pm$1.51
& 27.74$\pm$0.89 & 90.68$\pm$1.55
& 26.78$\pm$0.71 & 90.86$\pm$1.49
& 28.29$\pm$1.26 & 91.41$\pm$1.38
\\ \hline

\end{tabular}
\label{tab:ixi}
\end{table*}

\begin{figure*}
    \centering
    \includegraphics[width=0.9\textwidth]{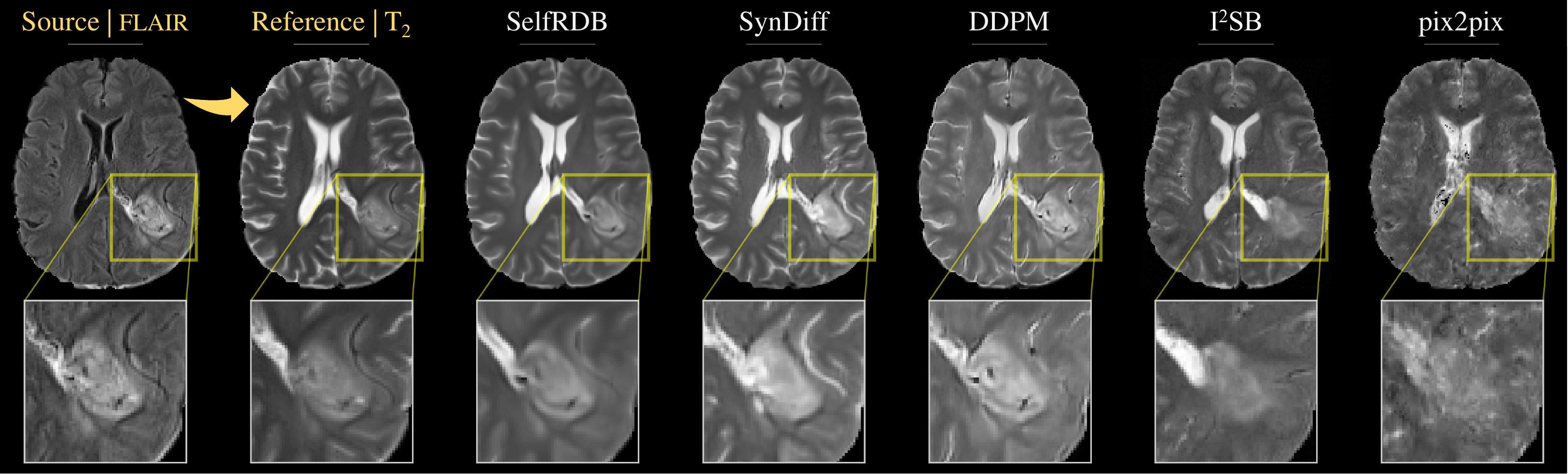}
    \caption{Multi-contrast MRI translation for a representative FLAIR$\rightarrow$T\textsubscript{2} task in the BRATS dataset. Synthesized target images for competing methods are shown along with the reference target image (i.e., ground truth) and the input source image.}
    \label{fig:brats_samples}
\end{figure*}

\begin{table*}[t]
\caption{Multi-contrast MRI translation in BRATS. PSNR (dB) and SSIM (\%) are listed as mean$\pm$std across the test set.}
\centering

\begin{tabular}{lcccccccc}
\hline
\multirow{2}{*}{} & \multicolumn{2}{c}{\TtwoTone} & \multicolumn{2}{c}{\ToneTtwo}  & \multicolumn{2}{c}{\FlairTtwo} & \multicolumn{2}{c}{\TtwoFlair} \\ 
\cline{2-9} & PSNR & SSIM & PSNR & SSIM & PSNR & SSIM & PSNR & SSIM \\ \hline

\multirow{1}{*}{SelfRDB} 
& \textbf{28.37$\pm$1.60} & \textbf{93.66$\pm$2.00}
& \textbf{27.42$\pm$2.19} & \textbf{92.95$\pm$2.86}
& \textbf{26.92$\pm$2.01} & \textbf{92.37$\pm$2.69}
& \textbf{28.06$\pm$1.83} & \textbf{90.70$\pm$2.63}
\\ \hline

\multirow{1}{*}{SynDiff} 
& 27.78$\pm$1.72 & 93.05$\pm$2.18
& 22.21$\pm$1.52 & 87.93$\pm$2.48
& 26.14$\pm$1.91 & 91.01$\pm$3.05
& 27.77$\pm$1.77 & 89.62$\pm$2.87
\\ \hline

\multirow{1}{*}{DDPM} 
& 27.47$\pm$1.28 & 92.24$\pm$1.99
& 25.97$\pm$2.09 & 90.24$\pm$3.41
& 25.40$\pm$1.76 & 89.79$\pm$3.01
& 26.90$\pm$1.84 & 87.86$\pm$2.85
\\ \hline

\multirow{1}{*}{I\textsuperscript{2}SB} 
& 22.24$\pm$2.18 & 79.87$\pm$5.84
& 21.80$\pm$2.33 & 80.75$\pm$5.49
& 23.28$\pm$2.33 & 84.38$\pm$4.34
& 25.92$\pm$2.00 & 83.51$\pm$4.05
\\ \hline

\multirow{1}{*}{pix2pix} 
& 27.75$\pm$1.26 & 91.66$\pm$2.44
& 26.37$\pm$2.07 & 91.56$\pm$2.81
& 24.46$\pm$2.14 & 86.10$\pm$3.99
& 26.86$\pm$1.76 & 86.95$\pm$3.41
\\ \hline

\end{tabular}
\label{tab:brats}
\end{table*}

\subsection{Competing Methods}
SelfRDB was demonstrated against state-of-the-art diffusion-based and adversarial methods. All competing methods were trained via supervised learning on paired source and target modalities. For each method, hyperparameter selection was performed to maximize performance on the validation set. A common set of parameters, including epochs, learning rate and loss-term weights, that attain near-optimal performance was selected across translation tasks. 

\subsubsection{SelfRDB} 
SelfRDB comprised generator and discriminator subnetworks. The generator was implemented with a residual UNet backbone with 12 residual stages equally split between encoding and decoding modules \cite{ronneberger2015u}. Each residual stage halved spatial resolution in the encoder, and doubled spatial resolution in the decoder module. Learnable time embeddings were computed via a multi-layer perceptron that received as input a 256-dimensional sinusoidal time encoding \cite{DDPM}. The time embeddings were added onto feature maps in each generator stage. The discriminator was implemented with a convolutional backbone with 6 stages \cite{adadiff}. Each stage halved spatial resolution, and time embeddings were also added onto feature maps in each discriminator stage. Cross-validated hyperparameters were 50 epochs, 10$^{-4}$ learning rate, $T$=1000, $\gamma$=2.2 , $\lambda_1$=1, $\lambda_2$=1. For recursive estimation of $\tilde{\boldsymbol{x}}_0^*$, convergence was assumed when the relative change in $\tilde{\boldsymbol{x}}_0^r$ between consecutive recursions fell below 1\%. 

\subsubsection{SynDiff}
An adversarial DDM model was considered with architecture, noise schedule and loss functions adopted from \cite{syndiff}. Cross-validated hyperparameters were 50 epochs, 15x10$^{-4}$ learning rate, $T$=1000, $k$=250 step size, adversarial loss weight of 1. 

\subsubsection{DDPM}
A DDM model was considered with architecture, noise schedule and loss functions adopted from \cite{nichol2021improved}. The source modality was input as stationary guidance to reverse diffusion steps. Cross-validated hyperparameters were 50 epochs, 10$^{-4}$ learning rate, $T$=1000. 

\subsubsection{I\textsuperscript{2}SB} 
A diffusion-bridge model was considered with architecture, noise schedule and loss functions adopted from \cite{liu2023i2sb}. The forward diffusion process mapped between source and target modalities. Cross-validated hyperparameters were 50 epochs, 10$^{-4}$ learning rate, $T$=1000. 

\subsubsection{pix2pix}
A GAN model was considered with architecture and loss functions adopted from \cite{dar2019image}. Cross-validated hyperparameters were 200 epochs, 2x10$^{-4}$ learning rate, and adversarial loss weight of 1.

\subsection{Modeling Procedures}
Models were implemented via the PyTorch framework and executed on Nvidia RTX 4090 GPUs. For training, Adam optimizer was used with $\beta_1$=0.5, $\beta_2$=0.9. For evaluation, a single target image was synthesized from the respective source image for each cross section. Model performance was assessed via peak signal-to-noise ratio (PSNR), and structural similarity index (SSIM) metrics. Prior to assessment, all images were normalized to a range of [0,1]. The significance of performance differences was examined via non-parametric Wilcoxon signed-rank tests (p$<$0.05).  
\section{Results}

\begin{figure*}
    \centering
    \includegraphics[width=\textwidth]{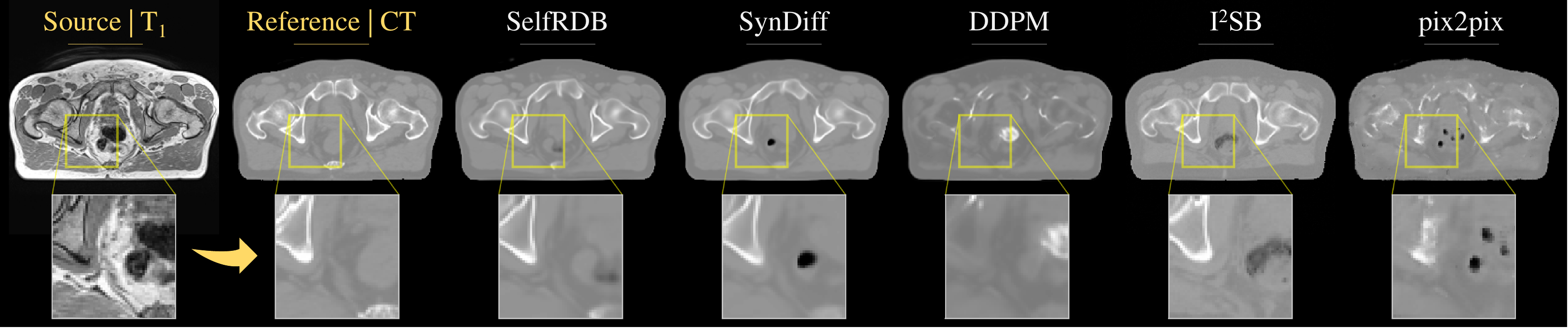}
    \caption{Multi-modal MRI-CT translation for a representative T\textsubscript{1}$\rightarrow$CT task in the pelvic dataset. Synthesized target images for competing methods are shown along with the reference target image (i.e., ground truth) and the input source image.}
    \label{fig:ct_samples}
\end{figure*}

\begin{table}[t]
\caption{Multi-modal MRI-CT translation in the pelvic dataset. PSNR (dB) and SSIM (\%) listed as mean$\pm$std across the test set.}
\centering

\begin{tabular}{lcccccccc}
\hline
\multirow{2}{*}{} & \multicolumn{2}{c}{\TtwoCT} & \multicolumn{2}{c}{\ToneCT} \\ 
\cline{2-5} & PSNR & SSIM & PSNR & SSIM \\ \hline

\multirow{1}{*}{SelfRDB} 
& \textbf{28.58$\pm$2.10} & \textbf{93.28$\pm$1.66}
& \textbf{27.86$\pm$3.37} & \textbf{92.69$\pm$5.28}
\\ \hline

\multirow{1}{*}{SynDiff}
& 26.54$\pm$2.01 & 89.59$\pm$2.61
& 27.41$\pm$4.68 & 92.07$\pm$5.32
\\ \hline

\multirow{1}{*}{DDPM}
& 26.88$\pm$1.96 & 91.18$\pm$2.03
& 26.39$\pm$2.54 & 90.62$\pm$5.04
\\ \hline

\multirow{1}{*}{I\textsuperscript{2}SB}
& 26.54$\pm$1.80 & 85.94$\pm$2.47
& 25.21$\pm$2.59 & 84.82$\pm$7.40
\\ \hline

\multirow{1}{*}{pix2pix} 
& 25.88$\pm$1.58 & 86.20$\pm$2.78
& 24.41$\pm$2.74 & 86.53$\pm$7.37
\\ \hline

\end{tabular}
\label{tab:ct}
\end{table}





\subsection{Multi-Contrast MRI Translation}
We first demonstrated SelfRDB in multi-contrast MRI translation tasks. The proposed method was compared against DDM models (SynDiff, DDPM), a diffusion bridge (I\textsuperscript{2}SB) and a GAN model (pix2pix). Evaluations were first conducted on the IXI dataset that contains brain images from healthy subjects. PSNR and SSIM metrics in IXI are listed in Table \ref{tab:ixi}. In each individual task, SelfRDB achieves significantly higher translation performance than all baselines (p$<$0.05). On average, SelfRDB outperforms DDMs by 1.25dB PSNR, 1.13\% SSIM, the diffusion bridge by 8.93dB PSNR, 24.77\% SSIM, and the GAN model by 3.62dB PSNR, 4.07\% SSIM. Representative target images synthesized by competing methods are displayed in Fig. \ref{fig:ixi_samples}. Among baselines, I\textsuperscript{2}SB shows generally poor anatomical fidelity to the target modality, and pix2pix suffers from prominent structural artifacts. Meanwhile, DDPM shows a degree of contrast loss and resultant blurring of tissue features, and SynDiff tends to over-flatten tissue signals that can yield loss of spatially-graded tissue features. In contrast, SelfRDB synthesizes target images with low artifact levels and a reliable depiction of fine tissue features.

We then evaluated competing methods on the BRATS dataset that contains brain images from glioma patients. PSNR and SSIM metrics in BRATS are listed in Table \ref{tab:brats}. Again, we find that SelfRDB achieves the highest translation performance among competing methods in each individual task (p$<$0.05). On average, SelfRDB outperforms DDMs by 1.49dB PSNR, 2.20\% SSIM, the diffusion bridge by 4.39dB PSNR, 10.29\% SSIM, and the GAN model by 1.34dB PSNR, 3.35\% SSIM. Representative target images synthesized by competing methods are displayed in Fig. \ref{fig:brats_samples}. Among baselines, I\textsuperscript{2}SB shows poor anatomical consistency, and pix2pix suffers from visible intensity artifacts, and DDPM shows a degree of spatial blurring. Meanwhile, SynDiff shows regions of gross intensity errors due to leakage of signal intensity and image artifacts from the source modality. In comparison, SelfRDB synthesizes target images with lower artifact levels and more accurate anatomical depiction.

\subsection{Multi-Modal MRI-CT Translation}
Next, we demonstrated SelfRDB in multi-modal MRI-CT translation tasks via comparisons against DDM models (SynDiff, DDPM), a diffusion bridge (I\textsuperscript{2}SB) and a GAN model (pix2pix). Evaluations were conducted on the pelvic MRI-CT dataset that contains healthy subjects. PSNR and SSIM metrics in the pelvic dataset are listed in Table \ref{tab:ct}. In each individual task, SelfRDB achieves significantly higher translation performance than all baselines (p$<$0.05). On average, SelfRDB outperforms DDMs by 1.42dB PSNR, 2.13\% SSIM, the diffusion bridge by 2.35dB PSNR, 7.61\% SSIM, and the GAN model by 3.08dB PSNR, 6.63\% SSIM. Representative target images synthesized by competing methods are displayed in Fig. \ref{fig:ct_samples}. Among baselines, DDPM and pix2pix suffer from gross intensity artifacts resulting in poor anatomical accuracy. While SynDiff and I\textsuperscript{2}SB are less amenable to these artifacts, they suffer from occasional leakage of signal intensities from the source modality. Instead, SelfRDB synthesizes target images with lower artifacts and higher anatomical accuracy.

\subsection{Ablation Studies}
We also conducted a systematic set of ablation studies to examine the contribution of main design elements in SelfRDB to translation performance. To assess the importance of the soft prior on the source modality, we formed a variant model with noise variance adopted from common diffusion bridges to attain the highest variance in mid-point of the diffusion process \cite{liu2023i2sb}. To assess the importance of stationary guidance from the original source image, we formed a variant model where the recovery network did not receive the original source image as an additional input but only received the noise-added source image at the end-point. To assess the importance of self-consistent target-image estimates, we formed a variant model that performed a one-shot estimation of $x_0$ in each reverse step \cite{DDPM}. Table \ref{tab:ablation_full} lists PSNR and SSIM metrics for SelfRDB and ablated variants on representative translation tasks. In all tasks, we find that SelfRDB outperforms ablated variants in translation performance (p$<$0.05). Taken together, these findings indicate that each proposed design element in SelfRDB makes an important contribution to its performance in multi-contrast MRI and multi-modal medical image translation. 

\begin{table}[t]
\caption{Performance of SelfRDB variants ablated of a soft prior on the source modality, of stationary source-image guidance, and of self-consistent target-image estimates.} 
\centering
\begin{tabular}{ccccccc}
\hline
\multirow{2}{*}{}          & \multicolumn{2}{c}{\TtwoTone (IXI)} & \multicolumn{2}{c}{\TtwoTone (BRATS)} & \multicolumn{2}{c}{\TtwoCT 
 (MRI-CT)} \\ \cline{2-7} 
                          & PSNR      & SSIM     & PSNR      & SSIM  & PSNR      & SSIM \\ \hline
\multirow{2}{*}{SelfRDB} 
    & \textbf{31.56} & \textbf{95.65}  &  \textbf{28.37} & \textbf{93.66}  &  \textbf{28.58 } & \textbf{93.28}      \\
    & \textbf{$\pm$1.58}  & \textbf{$\pm$1.18}  & \textbf{$\pm$1.60}  & \textbf{$\pm$2.00} & \textbf{$\pm$2.10}  & \textbf{$\pm$1.66} \\ \hline

\multirow{2}{*}{w/o soft prior} 
    &  30.16 & 94.54 & 27.54 & 92.01  &  28.35 & 92.73      \\
    &  $\pm$1.33 & $\pm$1.26  & $\pm$1.45  & $\pm$2.32  & $\pm$2.15  & $\pm$2.02 \\ \hline

\multirow{2}{*}{w/o source guidance} 
    &  20.43 & 75.39 & 21.10 & 77.92  &  24.70  & 83.53     \\
    &  $\pm$0.69 & $\pm$2.26  & $\pm$1.91  & $\pm$6.07  & $\pm$1.62  & $\pm$2.30 \\ \hline   

\multirow{2}{*}{w/o self-consistency }
    & 31.28 & 95.41  & 28.11 & 93.19  &  28.05 & 92.48     \\
    &  $\pm$1.5 & $\pm$1.16  & $\pm$1.65  & $\pm$2.14 &  $\pm$2.31 & $\pm$2.17 \\ \hline

\end{tabular}
\label{tab:ablation_full}
\end{table}
\section{Discussion}
SelfRDB is a diffusion bridge for medical image translation that progressively transforms a source onto a target modality. Compared to GANs amenable to training instabilities, it is a diffusion-based method that builds an explicit prior to improve sample fidelity. Compared to DDMs that are trained to learn a task-irrelevant noise-to-target transformation, it directly learns a source-to-target transformation of high task relevance. Compared to regular diffusion bridges, it leverages enhanced noise scheduling and sampling procedures to boost sampling accuracy. Our demonstrations on multi-modal translation tasks clearly suggest that these unique technical attributes of SelfRDB help significantly improve performance over baselines. 

The proposed method might benefit from several technical improvements to boost performance and efficiency in medical image translation. The first set of improvements concerns the reliability of translation models. SelfRDB draws intermediate image samples from a normal posterior probability similar to other diffusion-based methods, so it produces stochastic target images. Corroborating recent reports \cite{syndiff}, here we observed that multiple target images independently synthesized by SelfRDB show nominal variability (unreported). While this might be attributed to the diminishing noise variance towards the start-point of the diffusion process corresponding to the target modality, future research is warranted to evaluate the uncertainty of diffusion bridges in medical image translation. Note that, although SelfRDB is inherently a diffusion-based method, it employs an adversarial loss component that could induce susceptibility to training instabilities \cite{lan2020}. Here, we did not observe any notable sign of instabilities such as mode collapse or poor convergence when inspecting training and validation performance. Yet, when needed, spectral normalization or feature matching techniques could be adopted to improve training stability \cite{lan2020}. 

A second set of improvements concerns the selection of source-target modalities for the translation models. Here, we demonstrated high translation performance for SelfRDB when mapping between endogenous MRI contrasts (e.g., T\textsubscript{1}, T\textsubscript{2}), and between MRI and CT. Yet, there can be scenarios in which the primary tissue information needed to synthesize the target modality is only weakly present in the source modality. For instance, predicting exogenous MRI contrasts based on injection of external contrast agents from endogenous MRI contrasts \cite{lee2019}, or predicting MR images with enhanced soft-tissue differentiation from CT images with primarily bone and soft-tissue differentiation \cite{syndiff} are heavily ill-posed tasks. In such cases, translation performance might be improved by incorporating learned regularization terms regarding the target modality \cite{nie2018,ge2019,fedgimp}, or by including additional source modalities that carry a higher degree of correlated tissue information with the target modality \cite{mmgan,yurt2021mustgan}.

Lastly, a third set of improvements concerns the representational capacity and efficiency of model architectures. Here, supervised learning on paired source-target images within individual subjects was considered. However, it may not always be feasible to curate paired training sets of sufficient size for adequate model training. Cycle-consistent \cite{syndiff} or contrastive \cite{kim2024unpaired} architectures could be adopted to enable unsupervised learning for SelfRDB on unpaired data. Here, we employed a recovery network based on a convolutional backbone. Recent studies on medical imaging tasks report that transformer backbones can elevate sensitivity to long-range interactions \cite{resvit,TransGAN,transms} and enhance generalization performance to atypical anatomy \cite{slater,ssdiffrecon}. Adoption of a transformer backbone in SelfRDB could thus improve the representation of long-range context during source-to-target mapping. Note that SelfRDB has similar efficiency to conventional DDMs, so it has significantly longer inference times than GAN models that generate target images in a single forward pass. For efficiency improvements, acceleration approaches such as initiating sampling with an intermediate image derived from a secondary translation method \cite{chung2022cvpr}, or distillation of trained models onto fewer diffusion steps \cite{bedel2023dreamr} could be considered.
\section{Conclusion}
In this study, we introduced a novel diffusion bridge, SelfRDB, for multi-modal medical image translation tasks. SelfRDB learns a task-relevant progressive transformation between source- and target-modality distributions. In reverse diffusion steps, it improves sample reliability via a self-consistent recursive sampling procedure based on conditional guidance from the acquired source image. It further employs a monotonically-increasing scheduling for the noise variance towards the source image in order to facilitate information transfer between the modalities. SelfRDB achieves superior image quality over state-of-the-art GAN and diffusion methods, so it holds great promise for medical image translation applications.

\bibliographystyle{IEEETran} 
\bibliography{IEEEabrv,refs}

\begin{thebibliography}{10}
\providecommand{\url}[1]{#1}
\csname url@samestyle\endcsname
\providecommand{\newblock}{\relax}
\providecommand{\bibinfo}[2]{#2}
\providecommand{\BIBentrySTDinterwordspacing}{\spaceskip=0pt\relax}
\providecommand{\BIBentryALTinterwordstretchfactor}{4}
\providecommand{\BIBentryALTinterwordspacing}{\spaceskip=\fontdimen2\font plus
\BIBentryALTinterwordstretchfactor\fontdimen3\font minus \fontdimen4\font\relax}
\providecommand{\BIBforeignlanguage}[2]{{%
\expandafter\ifx\csname l@#1\endcsname\relax
\typeout{** WARNING: IEEEtran.bst: No hyphenation pattern has been}%
\typeout{** loaded for the language `#1'. Using the pattern for}%
\typeout{** the default language instead.}%
\else
\language=\csname l@#1\endcsname
\fi
#2}}
\providecommand{\BIBdecl}{\relax}
\BIBdecl

\bibitem{iglesias2013}
J.~E. Iglesias, E.~Konukoglu, D.~Zikic, B.~Glocker, K.~Van~Leemput, and B.~Fischl, ``Is synthesizing \uppercase{mri} contrast useful for inter-modality analysis?'' in \emph{Med Image Comput Comput Assist Interv}, 2013, pp. 631--638.

\bibitem{lee2017}
J.~Lee, A.~Carass, A.~Jog, C.~Zhao, and J.~Prince, ``Multi-atlas-based \uppercase{ct} synthesis from conventional \uppercase{mri} with patch-based refinement for \uppercase{mri}-based radiotherapy planning,'' in \emph{SPIE Med Imag.}, vol. 10133, 2017, p. 101331I.

\bibitem{ye2013}
D.~H. Ye, D.~Zikic, B.~Glocker, A.~Criminisi, and E.~Konukoglu, ``Modality propagation: Coherent synthesis of subject-specific scans with data-driven regularization,'' in \emph{Med Image Comput Comput Assist Interv}, 2013, pp. 606--613.

\bibitem{huynh2015}
T.~Huynh, Y.~Gao, J.~Kang, L.~Wang, P.~Zhang, J.~Lian, and D.~Shen, ``Estimating \uppercase{ct} image from \uppercase{mri} data using structured random forest and auto-context model,'' \emph{{IEEE} Trans Med Imag}, vol.~35, no.~1, pp. 174--183, 2016.

\bibitem{jog2017}
A.~Jog, A.~Carass, S.~Roy, D.~L. Pham, and J.~L. Prince, ``Random forest regression for magnetic resonance image synthesis,'' \emph{Med Image Anal}, vol.~35, pp. 475--488, 2017.

\bibitem{joyce2017}
T.~Joyce, A.~Chartsias, and S.~A. Tsaftaris, ``Robust multi-modal \uppercase{mr} image synthesis,'' in \emph{Med Image Comput Comput Assist Interv}, 2017, pp. 347--355.

\bibitem{cordier2016}
N.~Cordier, H.~Delingette, M.~Le, and N.~Ayache, ``Extended modality propagation: Image synthesis of pathological cases,'' \emph{{IEEE} Trans Med Imag}, vol.~35, pp. 2598--2608, 2016.

\bibitem{wu2016}
Y.~Wu, W.~Yang, L.~Lu, Z.~Lu, L.~Zhong, M.~Huang, Y.~Feng, Q.~Feng, and W.~Chen, ``Prediction of \uppercase{ct} substitutes from \uppercase{mr} images based on local diffeomorphic mapping for brain \uppercase{pet} attenuation correction,'' \emph{J Nucl Med}, vol.~57, no.~10, pp. 1635--1641, 2016.

\bibitem{zhao2017}
C.~Zhao, A.~Carass, J.~Lee, Y.~He, and J.~L. Prince, ``Whole brain segmentation and labeling from \uppercase{ct} using synthetic \uppercase{mr} images,'' in \emph{Mach Learn Med Imaging}, 2017, pp. 291--298.

\bibitem{huang2018}
Y.~Huang, L.~Shao, and A.~F. Frangi, ``Cross-modality image synthesis via weakly coupled and geometry co-regularized joint dictionary learning,'' \emph{{IEEE} Trans Med Imag}, vol.~37, no.~3, pp. 815--827, 2018.

\bibitem{lee2019}
D.~Lee, J.~Kim, W.-J. Moon, and J.~C. Ye, ``Colla\uppercase{gan}: Collaborative \uppercase{gan} for missing image data imputation,'' in \emph{Comput Vis Pattern Recognit}, 2019, pp. 2487--2496.

\bibitem{divbar2019}
L.~T. Clark \emph{et~al.}, ``{Increasing Diversity in Clinical Trials: Overcoming Critical Barriers},'' \emph{Cur. Prob. Cardiol.}, vol.~44, no.~5, pp. 148--172, 2019.

\bibitem{roy2013}
S.~Roy, A.~Jog, A.~Carass, and J.~L. Prince, ``Atlas based intensity transformation of brain \uppercase{mr} images,'' in \emph{Multimodal Brain Image Anal.}, 2013, pp. 51--62.

\bibitem{alexander2014}
D.~C. Alexander, D.~Zikic, J.~Zhang, H.~Zhang, and A.~Criminisi, ``Image quality transfer via random forest regression: Applications in diffusion \uppercase{mri},'' in \emph{Med Image Comput Comput Assist Interv}, 2014, pp. 225--232.

\bibitem{huang2017}
Y.~Huang, L.~Shao, and A.~F. Frangi, ``Simultaneous super-resolution and cross-modality synthesis of {3D} medical images using weakly-supervised joint convolutional sparse coding,'' \emph{Comput Vis Pattern Recognit}, pp. 5787--5796, 2017.

\bibitem{hien2015}
H.~Van~Nguyen, K.~Zhou, and R.~Vemulapalli, ``Cross-domain synthesis of medical images using efficient location-sensitive deep network,'' in \emph{Med Image Comput Comput Assist Interv}, 2015, pp. 677--684.

\bibitem{vemulapalli2015}
R.~Vemulapalli, H.~Van~Nguyen, and S.~K. Zhou, ``Unsupervised cross-modal synthesis of subject-specific scans,'' in \emph{Int Conf Comput Vis}, 2015, pp. 630--638.

\bibitem{sevetlidis2016}
V.~Sevetlidis, M.~V. Giuffrida, and S.~A. Tsaftaris, ``Whole image synthesis using a deep encoder-decoder network,'' in \emph{Simul Synth Med Imaging}, 2016, pp. 127--137.

\bibitem{nie2016}
D.~Nie, X.~Cao, Y.~Gao, L.~Wang, and D.~Shen, ``Estimating \uppercase{ct} image from \uppercase{mri} data using {3D} fully convolutional networks,'' in \emph{Deep Learn Data Label Med Appl}, 2016, pp. 170--178.

\bibitem{bowles2016}
C.~Bowles \emph{et~al.}, ``Pseudo-healthy image synthesis for white matter lesion segmentation,'' in \emph{Simul Synth Med Imaging}, 2016, pp. 87--96.

\bibitem{chartsias2018}
A.~Chartsias, T.~Joyce, M.~V. Giuffrida, and S.~A. Tsaftaris, ``Multimodal \uppercase{mr} synthesis via modality-invariant latent representation,'' \emph{{IEEE} Trans Med Imag}, vol.~37, no.~3, pp. 803--814, 2018.

\bibitem{wei2019}
W.~Wei, E.~Poirion, B.~Bodini, S.~Durrleman, O.~Colliot, B.~Stankoff, and N.~Ayache, ``Fluid-attenuated inversion recovery \uppercase{mri} synthesis from multisequence \uppercase{mri} using three-dimensional fully convolutional networks for multiple sclerosis,'' \emph{J Med Imaging}, vol.~6, no.~1, p. 014005, 2019.

\bibitem{yu2018}
B.~Yu, L.~Zhou, L.~Wang, J.~Fripp, and P.~Bourgeat, ``{3D} c\uppercase{gan} based cross-modality \uppercase{mr} image synthesis for brain tumor segmentation,'' \emph{Int. Symp. Biomed. Imaging}, pp. 626--630, 2018.

\bibitem{armanious2019}
K.~Armanious, C.~Jiang, M.~Fischer, T.~Küstner, T.~Hepp, K.~Nikolaou, S.~Gatidis, and B.~Yang, ``Med\uppercase{gan}: Medical image translation using \uppercase{gan}s,'' \emph{Comput Med Imaging Grap}, vol.~79, p. 101684, 2019.

\bibitem{li2019}
H.~Li, J.~C. Paetzold, A.~Sekuboyina, F.~Kofler, J.~Zhang, J.~S. Kirschke, B.~Wiestler, and B.~Menze, ``{DiamondGAN: Unified} multi-modal generative adversarial networks for \uppercase{mri} sequences synthesis,'' in \emph{Med. Image Comput Comput Assist Interv}, 2019, pp. 795--803.

\bibitem{nie2018}
D.~Nie, R.~Trullo, J.~Lian, L.~Wang, C.~Petitjean, S.~Ruan, and Q.~Wang, ``Medical image synthesis with deep convolutional adversarial networks,'' \emph{{IEEE} Trans. Biomed. Eng.}, vol.~65, no.~12, pp. 2720--2730, 2018.

\bibitem{dar2019image}
S.~U. Dar, M.~Yurt, L.~Karacan, A.~Erdem, E.~Erdem, and T.~Cukur, ``Image synthesis in multi-contrast {MRI} with conditional generative adversarial networks,'' \emph{IEEE Trans Med Imaging}, vol.~38, no.~10, pp. 2375--2388, 2019.

\bibitem{yu2019}
B.~Yu, L.~Zhou, L.~Wang, Y.~Shi, J.~Fripp, and P.~Bourgeat, ``{Ea}-\uppercase{gan}s: {Edge}-aware generative adversarial networks for cross-modality \uppercase{mr} image synthesis,'' \emph{{IEEE} Trans Med Imag}, vol.~38, no.~7, pp. 1750--1762, 2019.

\bibitem{yang2018}
H.~Yang, J.~Sun, A.~Carass, C.~Zhao, J.~Lee, Z.~Xu, and J.~Prince, ``Unpaired brain \uppercase{mr}-to-\uppercase{ct} synthesis using a structure-constrained cycle\uppercase{gan},'' \emph{arXiv:1809.04536}, 2018.

\bibitem{jin2018}
C.-B. Jin, H.~Kim, M.~Liu, W.~Jung, S.~Joo, E.~Park, Y.~S. Ahn, I.~H. Han, J.~I. Lee, and X.~Cui, ``Deep \uppercase{ct} to \uppercase{mr} synthesis using paired and unpaired data,'' \emph{Sensors}, vol.~19, no.~10, p. 2361, 2019.

\bibitem{resvit}
O.~Dalmaz, M.~Yurt, and T.~Çukur, ``{ResViT: Residual} vision transformers for multi-modal medical image synthesis,'' \emph{IEEE Trans Med Imaging}, vol.~44, no.~10, pp. 2598--2614, 2022.

\bibitem{wang2020}
G.~Wang, E.~Gong, S.~Banerjee, D.~Martin, E.~Tong, J.~Choi, H.~Chen, M.~Wintermark, J.~M. Pauly, and G.~Zaharchuk, ``Synthesize high-quality multi-contrast magnetic resonance imaging from multi-echo acquisition using multi-task deep generative model,'' \emph{{IEEE} Trans Med Imag}, vol.~39, no.~10, pp. 3089--3099, 2020.

\bibitem{zhou2020}
T.~Zhou, H.~Fu, G.~Chen, J.~Shen, and L.~Shao, ``{Hi-Net: H}ybrid-fusion network for multi-modal \uppercase{mr} image synthesis,'' \emph{{IEEE} Trans Med Imag}, vol.~39, no.~9, pp. 2772--2781, 2020.

\bibitem{syndiff}
M.~{\"O}zbey, S.~U. Dar, H.~A. Bedel, O.~Dalmaz, {\c{S}}.~{\"O}zturk, A.~G{\"u}ng{\"o}r, and T.~{\c{C}}ukur, ``Unsupervised medical image translation with adversarial diffusion models,'' \emph{IEEE Trans Med Imaging}, vol.~42, no.~12, pp. 3524--3539, 2023.

\bibitem{meng_arxiv_2022}
X.~Meng, Y.~Gu, Y.~Pan, N.~Wang, P.~Xue, M.~Lu, X.~He, Y.~Zhan, and D.~Shen, ``A novel unified conditional score-based generative framework for multi-modal medical image completion,'' \emph{arXiv:2207.03430}, 2022.

\bibitem{lyu_arxiv_2022}
Q.~Lyu and G.~Wang, ``Conversion between {CT and MRI} images using diffusion and score-matching models,'' \emph{arXiv:2209.12104}, 2022.

\bibitem{wang2024tmi}
Z.~Wang, Y.~Yang, Y.~Chen, T.~Yuan, M.~Sermesant, H.~Delingette, and O.~Wu, ``Mutual information guided diffusion for zero-shot cross-modality medical image translation,'' \emph{IEEE Trans Med Imaging}, pp. 1--1, 2024.

\bibitem{song2021solving}
Y.~Song, L.~Shen, L.~Xing, and S.~Ermon, ``Solving inverse problems in medical imaging with score-based generative models,'' \emph{arXiv:2111.08005}, 2021.

\bibitem{adadiff}
A.~Güngör, S.~U. Dar, Şaban Öztürk, Y.~Korkmaz, H.~A. Bedel, G.~Elmas, M.~Ozbey, and T.~Çukur, ``Adaptive diffusion priors for accelerated mri reconstruction,'' \emph{Med Image Anal}, p. 102872, 2023.

\bibitem{DDPM}
J.~Ho, A.~Jain, and P.~Abbeel, ``Denoising diffusion probabilistic models,'' in \emph{Adv Neural Inf Process Syst}, vol.~33, 2020, pp. 6840--6851.

\bibitem{delbracio2023inversion}
M.~Delbracio and P.~Milanfar, ``Inversion by direct iteration: An alternative to denoising diffusion for image restoration,'' \emph{Tran. Mach. Learn. Res.}, 2023.

\bibitem{chung2023direct}
H.~Chung, J.~Kim, and J.~C. Ye, ``Direct diffusion bridge using data consistency for inverse problems,'' \emph{arXiv:2305.19809}, 2023.

\bibitem{liu2023i2sb}
G.-H. Liu, A.~Vahdat, D.-A. Huang, E.~A. Theodorou, W.~Nie, and A.~Anandkumar, ``{I$^2$SB: Image-to-Image Schr\"odinger Bridge},'' \emph{arXiv:2302.05872}, 2023.

\bibitem{kim2024unpaired}
B.~Kim, G.~Kwon, K.~Kim, and J.~C. Ye, ``Unpaired image-to-image translation via neural schr\"odinger bridge,'' in \emph{ICLR}, 2024.

\bibitem{FDB}
M.~U. Mirza, O.~Dalmaz, H.~A. Bedel, G.~Elmas, Y.~Korkmaz, A.~Gungor, S.~U. Dar, and T.~Çukur, ``{Learning Fourier-Constrained Diffusion Bridges for MRI Reconstruction},'' \emph{arXiv:2308.01096}, 2023.

\bibitem{kim2024hicbridge}
\BIBentryALTinterwordspacing
J.~Kim and J.~C. Ye, ``Hi{CB}ridge: Resolution enhancement of hi-c data using direct diffusion bridge,'' 2024. [Online]. Available: \url{https://openreview.net/forum?id=RUvzlotXY0}
\BIBentrySTDinterwordspacing

\bibitem{DDIB}
X.~Su, J.~Song, C.~Meng, and S.~Ermon, ``Dual diffusion implicit bridges for image-to-image translation,'' \emph{arXic:2203.08382}, 2023.

\bibitem{peng2022}
C.~Peng, P.~Guo, S.~K. Zhou, V.~Patel, and R.~Chellappa, ``Towards performant and reliable undersampled mr reconstruction via diffusion model sampling,'' \emph{arXiv:2203.04292}, 2022.

\bibitem{wolterink2017}
J.~M. Wolterink, A.~M. Dinkla, M.~H.~F. Savenije, P.~R. Seevinck, C.~A.~T. van~den Berg, and I.~I{\v{s}}gum, ``Deep {MR} to {CT} synthesis using unpaired data,'' in \emph{Simul Synth Med Imaging}, Cham, 2017, pp. 14--23.

\bibitem{dong2019}
X.~Dong, T.~Wang, Y.~Lei, K.~Higgins, T.~Liu, W.~Curran, H.~Mao, J.~Nye, and X.~Yang, ``Synthetic \uppercase{ct} generation from non-attenuation corrected \uppercase{pet} images for whole-body \uppercase{pet} imaging,'' \emph{Phys Med Biol}, vol.~64, no.~21, p. 215016, 2019.

\bibitem{daras2022soft}
G.~Daras, M.~Delbracio, H.~Talebi, A.~G. Dimakis, and P.~Milanfar, ``Soft diffusion: Score matching for general corruptions,'' \emph{arXiv:2209.05442}, 2022.

\bibitem{chen2021likelihood}
T.~Chen, G.-H. Liu, and E.~A. Theodorou, ``Likelihood training of schr$\backslash$" odinger bridge using forward-backward sdes theory,'' \emph{arXiv preprint arXiv:2110.11291}, 2021.

\bibitem{nelson1967}
E.~Nelson, \emph{Dynamical Theories of Brownian Motion}.\hskip 1em plus 0.5em minus 0.4em\relax Princeton University Press, 1967.

\bibitem{song2020score}
Y.~Song, J.~Sohl-Dickstein, D.~P. Kingma, A.~Kumar, S.~Ermon, and B.~Poole, ``Score-based generative modeling through stochastic differential equations,'' \emph{arXiv:2011.13456}, 2020.

\bibitem{chen2023schrodinger}
Z.~Chen, G.~He, K.~Zheng, X.~Tan, and J.~Zhu, ``Schrodinger bridges beat diffusion models on text-to-speech synthesis,'' \emph{arXiv preprint arXiv:2312.03491}, 2023.

\bibitem{song2019generative}
Y.~Song and S.~Ermon, ``Generative modeling by estimating gradients of the data distribution,'' \emph{Advances in neural information processing systems}, vol.~32, 2019.

\bibitem{DiffNvidia}
Z.~Xiao, K.~Kreis, and A.~Vahdat, ``Tackling the generative learning trilemma with denoising diffusion {GAN}s,'' in \emph{Int Conf Learn Represent}, 2022.

\bibitem{pgan}
S.~U. Dar, M.~Yurt, L.~Karacan, A.~Erdem, E.~Erdem, and T.~Çukur, ``Image synthesis in multi-contrast \uppercase{mri} with conditional generative adversarial networks,'' \emph{{IEEE} Trans Med Imag}, vol.~38, no.~10, pp. 2375--2388, 2019.

\bibitem{brats_1}
B.~H. Menze \emph{et~al.}, ``The multimodal brain tumor image segmentation benchmark {(BRATS)},'' \emph{{IEEE} Trans Med Imag}, vol.~34, no.~10, pp. 1993--2024, 2015.

\bibitem{mr_ct_dataset}
T.~Nyholm \emph{et~al.}, ``\uppercase{mr} and \uppercase{ct} data with multiobserver delineations of organs in the pelvic area—part of the gold atlas project,'' \emph{Med Phys}, vol.~45, no.~3, pp. 1295--1300, 2018.

\bibitem{fslcitation}
M.~Jenkinson and S.~Smith, ``A global optimisation method for robust affine registration of brain images,'' \emph{Med Image Anal}, vol.~5, pp. 143--156, 2001.

\bibitem{ronneberger2015u}
O.~Ronneberger, P.~Fischer, and T.~Brox, ``{U-net: C}onvolutional networks for biomedical image segmentation,'' in \emph{Med Image Comput Comput Assist Inter}.\hskip 1em plus 0.5em minus 0.4em\relax Springer, 2015, pp. 234--241.

\bibitem{nichol2021improved}
A.~Q. Nichol and P.~Dhariwal, ``Improved denoising diffusion probabilistic models,'' in \emph{Int Conf Mach Learn}, 2021, pp. 8162--8171.

\bibitem{lan2020}
H.~Lan, A.~Toga, and F.~Sepehrband, ``{SC-GAN: 3D} self-attention conditional \uppercase{gan} with spectral normalization for multi-modal neuroimaging synthesis,'' \emph{bioRxiv:2020.06.09.143297}, 2020.

\bibitem{ge2019}
Y.~Ge, D.~Wei, Z.~Xue, Q.~Wang, X.~Zhou, Y.~Zhan, and S.~Liao, ``Unpaired \uppercase{mr} to \uppercase{ct} synthesis with explicit structural constrained adversarial learning,'' in \emph{Int. Symp. Biomed. Imaging}, 2019, pp. 1096--1099.

\bibitem{fedgimp}
G.~Elmas, S.~U. Dar, Y.~Korkmaz, E.~Ceyani, B.~Susam, M.~Özbey, S.~Avestimehr, and T.~Çukur, ``{Federated Learning of Generative Image Priors for MRI Reconstruction},'' \emph{IEEE Trans Med Imaging}, vol.~42, no.~7, pp. 1996--2009, 2023.

\bibitem{mmgan}
A.~Sharma and G.~Hamarneh, ``Missing \uppercase{mri} pulse sequence synthesis using multi-modal generative adversarial network,'' \emph{{IEEE} Trans Med Imag}, vol.~39, pp. 1170--1183, 2020.

\bibitem{yurt2021mustgan}
M.~Yurt, S.~U. Dar, A.~Erdem, E.~Erdem, K.~K. Oguz, and T.~Çukur, ``must\uppercase{gan}: multi-stream generative adversarial networks for \uppercase{mr} image synthesis,'' \emph{Med Image Anal}, vol.~70, p. 101944, 2021.

\bibitem{TransGAN}
Y.~Luo, Y.~Wang, C.~Zu, B.~Zhan, X.~Wu, J.~Zhou, D.~Shen, and L.~Zhou, ``{3D Transformer-GAN} for high-quality {PET} reconstruction,'' in \emph{Med Image Comput Comput Assist Interv}, 2021, pp. 276--285.

\bibitem{transms}
A.~Gungor, B.~Askin, D.~A. Soydan, E.~U. Saritas, C.~B. Top, and T.~Çukur, ``{TranSMS: Transformers} for super-resolution calibration in magnetic particle imaging,'' \emph{IEEE Trans Med Imaging}, vol.~41, no.~12, pp. 3562--3574, 2022.

\bibitem{slater}
Y.~Korkmaz, S.~U.~H. Dar, M.~Yurt, M.~Ozbey, and T.~Cukur, ``Unsupervised {MRI} reconstruction via zero-shot learned adversarial transformers,'' \emph{IEEE Trans Med Imaging}, vol.~41, no.~7, pp. 1747--1763, 2022.

\bibitem{ssdiffrecon}
Y.~Korkmaz, T.~Cukur, and V.~M. Patel, ``Self-supervised mri reconstruction with unrolled diffusion models,'' in \emph{MICCAI}, 2023, pp. 491--501.

\bibitem{chung2022cvpr}
H.~Chung, B.~Sim, and J.~C. Ye, ``Come-closer-diffuse-faster: Accelerating conditional diffusion models for inverse problems through stochastic contraction,'' in \emph{IEEE Conf Comput Vis Pattern Recognit}, 2022, pp. 12\,413--12\,422.

\bibitem{bedel2023dreamr}
H.~A. Bedel and T.~Çukur, ``{DreaMR: Diffusion-driven} counterfactual explanation for functional {MRI},'' \emph{arXiv:2307.09547}, 2023.

\end{thebibliography}

\end{document}